\documentclass[sigconf,authorversion,nonacm]{acmart}
\AtBeginDocument{%
  }

\setcopyright{acmlicensed}
\copyrightyear{2018}
\acmYear{2018}
\acmDOI{XXXXXXX.XXXXXXX}
\acmConference[Conference acronym 'XX]{Make sure to enter the correct
  conference title from your rights confirmation email}{June 03--05,
  2018}{Woodstock, NY}

\usepackage{bm}
\usepackage{braket}
\usepackage{todonotes}
\usepackage{subcaption}
\usepackage{listings}
\usepackage{multirow}
\newcommand{\Sdg}{S\textsuperscript{\dag}}

\usepackage{color}
\usepackage{ulem}

\begin{document}

\title{Multilevel Circuit Optimization in Quantum Compilers: \\ A Case Study}

\author{Tamiya Onodera}
\affiliation{%
  \institution{IBM Quantum Japan}
  \city{Tokyo}
  \country{Japan}
}
\affiliation{%
  \institution{Quantum Computing Center, Keio University}
  \city{Kanagawa}
  \country{Japan}
}

\author{Yuki Sato}
\affiliation{%
  \institution{Toyota Central R\&D Labs., Inc.}
  \city{Tokyo}
  \country{Japan}
}
\affiliation{%
  \institution{Quantum Computing Center, Keio University}
  \city{Kanagawa}
  \country{Japan}
}

\author{Toshinari Itoko}
\affiliation{%
  \institution{IBM Quantum Japan}
  \city{Tokyo}
  \country{Japan}
}
\affiliation{%
  \institution{Quantum Computing Center, Keio University}
  \city{Kanagawa}
  \country{Japan}
}

\author{Naoki Yamamoto}
\affiliation{%
  \institution{Department of Applied Physics and Physico-Informatics, Keio University}
  \city{Kanagawa}
  \country{Japan}
}
\affiliation{%
  \institution{Quantum Computing Center, Keio University}
  \city{Kanagawa}
  \country{Japan}
}

\renewcommand{\shortauthors}{Onodera et al.}

\begin{abstract}
In this paper,  we explore {\it multilevel circuit optimization} (MLCO), 
where we deploy multiple gate sets and progressively lower the source circuit through the gate sets to the target circuit. At each level, we first perform an appropriate set of circuit simplifications
and then lower the simplified circuit into the next level, decomposing the gates not supported there. We demonstrate its effectiveness, using as a case study the source circuit for Hamiltonian simulation to solve a partial differential equation, which is densely populated with multi-controlled gates and is transformed by the state-of-the-art circuit compiler to the target circuit with the quadratic number of CX gates in the number of qubits. 
MLCO makes visible higher-level circuit structures, 
providing us with insights about how to simplify the circuits and how to decompose the gates.
By putting the right circuit structure in place
and selecting the right decomposition algorithm,  
we could cause massive cancellation of entangling gates,
thereby having achieved the quadratic reduction in the number of CX gates.
\end{abstract}

\keywords{Circuit optimization, Multi-controlled gates, Hamiltonian simulation, Partial differential equation}

\maketitle

\section{Introduction}
There has been a growing amount of  literature on solving problems with quantum computers. 
One of the most fundamental quantum algorithms is the Hamiltonian simulation~\cite{nielsen2010quantum, berry2015simulating, bosse2025efficient}, which is used to prepare states evolved by a Hamiltonian; this also functions as a subroutine of many quantum algorithms, including quantum phase estimation~\cite{nielsen2010quantum, wiebe2016efficient, lin2022heisenberg} and linear system solvers~\cite{harrow2009quantum, childs2017quantum}. 
In Hamiltonian simulation, we encode a problem of interest to a Hamiltonian and perform circuit synthesis of its time evolution to create the {\it source circuit}, which 
is then optimized and compiled into the {\it target circuit} that conforms to the hardware gate set for a quantum computer. Typically, the hardware gate set consists of a few single-qubit gates and one or more two-qubit gates~\cite{li2023qasmbench}.

Recently, Sato et al. presented a method to build the source circuit 
for Hamiltonian simulation to solve a partial differential equation (PDE)
such as advection and wave equations~\cite{sato2024hamiltonian}. 
While the work represents a significant departure 
from prior art~\cite{babbush2023exponential, jin2023aquantum, an2023linear}, in which only oracle-based circuits had been shown,
the source circuits built to the method are densely populated with multi-controlled gates
(See Figure~\ref{fig:source} for two examples).
They estimated that the $n$-qubit source circuit
is compiled to the target circuit with the number of CX (controlled X, also known as CNOT) gates quadratic in $n$,
which makes even a small experiment unlikely to run on a near-term noisy quantum computer.

We actually optimized the PDE circuits with the state-of-the-art transpiler\footnote{
Qiskit uses the word “transpiler” for what others might call a compiler,
to emphasize its nature as a circuit-to-circuit transformation tool.} in Qiskit~\cite{qiskit2024}, which possesses powerful optimization capabilities such as gate cancellation~\cite{nam2018automated} and template optimization~\cite{maslov2008quantum,iten2022template}. 
Yet it still resulted in the target circuits with quadratic numbers of CX gates in $n$.  As we will discuss, this is primarily because the transpiler decomposes the PDE circuit at an early stage into single- and two-qubit gates and most optimization opportunities for the PDE circuit are lost at that low level.

In this paper, we explore {\it Multilevel Circuit Optimization} (MLCO) and demonstrate its effectiveness, using the PDE circuit as a case study.
In MLCO, we deploy multiple gate sets and progressively lower the source circuit through the gate sets to the target circuit. 
At each level, we first perform an appropriate set of circuit simplifications and then lower the simplified circuit into the next level, decomposing the gates supported at the next level.
Note that MLCO follows the tradition of a classical compiler,
where they deploy multilevel IRs (Intermediate Representations) to allow optimizations to be implemented at the IR levels best suited for them~\cite{stanier2013IRs}.

Having a clear picture of a circuit at a high level provides us with insights about how to simplify the circuit and how to decompose the gates.
As we will show, by putting the right circuit structure in place
and selecting the right decomposition algorithm,  
we can cause massive cancellation of entangling gates,
thereby having achieved the quadratic reduction in the number of CX gates.

The remainder of the paper is organized as follows.
Section 2 describes Hamiltonian simulation for a PDE, showing how source circuits are derived from their Hamiltonian.
Section 3 explains MLCO, using the PDE circuit as a case study.  Section 4 shows its implementation, while Section 5 presents results.
Section 6 discusses related work.
Finally, Section 7 concludes the paper.

\section{Hamiltonian simulation for Partial Differential Equations}
\label{sec:PDE}
We sketch out how we create a Hamiltonian to solve a PDE and how we build a circuit from the constructed Hamiltonian. 
For the full details, refer to~\cite{sato2024hamiltonian}.

\subsection{Finite difference operator}
Let us consider a one-dimensional domain $\Omega := (0, L)$.
Discretizing it into an equally spaced grid with $N+2=2^n+2$ nodes and $l := L/(N+1)$ intervals, we represent the scalar function $\psi(x)$ for $x \in \Omega$ by the value evaluated at each node as $\bm{\psi} := [\psi_{-1}, \psi_0, \dots, \psi_{N-1}, \psi_{N}]$ where $\psi_j$ for $0 \leq j \leq N-1$ is the value of the scalar field at the $j$-th interior node and $\psi_{-1}$ and $\psi_{N}$ are those on the boundary nodes.
Since the values on the boundary node are determined by boundary conditions, we focus on the values on the interior nodes, expressing them as a quantum state vector:
\begin{align}
    \ket{\psi} := \frac{1}{\| \psi \|} \sum_{j=0}^{2^n-1} \psi_j \ket{j},
\end{align}
where $\| \psi \| := \sqrt{\sum_{j=0}^{2^n-1} \psi_j^2}$ and $\ket{j} := \ket{j_{n-1} \dots j_0}$ with $j_{\cdot} \in \{0, 1 \}$ is a computational basis.
To obtain finite difference operators acting on the quantum state vector, we consider shift operators $S^{-}$ and $S^{+}$ for the computational basis, which can be expressed using ladder operators $\sigma_{01} := \ket{0} \! \bra{1}$ and $\sigma_{10} := \ket{1} \! \bra{0}$, as follows~\cite{sato2024hamiltonian}:
\begin{align}
    S^{-} := \sum_{j=1}^{2^n-1} \ket{j-1} \! \bra{j} = \sum_{j=1}^n I^{\otimes (n-j)} \otimes \sigma_{01} \otimes \sigma_{10}^{\otimes (j-1)} \\
    S^{+} := \sum_{j=1}^{2^n-1} \ket{j} \! \bra{j-1} = \sum_{j=1}^n I^{\otimes (n-j)} \otimes \sigma_{10} \otimes \sigma_{01}^{\otimes (j-1)},
\end{align}
where $I$ is an identity operator.
Finite difference operators can be represented using the shift operators, yielding Hamiltonians for PDEs~\cite{sato2025quantum}.
For instance, the forward, backward, and central differential operators with the Dirichlet boundary condition are respectively given as
\begin{align}
    D^{+} &= \frac{1}{l} \left( S^{-} - I^{\otimes n} \right), \\
    D^{-} &= \frac{1}{l} \left( I^{\otimes n} - S^{+} \right), \\
    D^{\pm} &= \frac{1}{2l} \left( S^{-} - S^{+} \right).
\end{align}

\subsection{Hamiltonian derived from finite difference operators}
This study focuses on a one-dimensional wave equation:
\begin{align}
    \frac{\partial^2 u(t, x)}{\partial t^2} = c^2 \frac{\partial^2 u(t, x)}{\partial x^2},
\end{align}
where $u(t, x)$ is the state governed by the wave equation (e.g., displacement or pressure), $c$ is the speed of the wave, $t$ is the time, $x$ is the spatial coordinate.
Discretizing the spatial derivative using the finite difference operators, the corresponding $(n+1)$-qubit Hamiltonian reads
\begin{align}
    H &:= c( \sigma_{01} \otimes D^{+} - \sigma_{10} \otimes D^{-}),
\end{align}
Note that $H$ is Hermitian, i.e., $H^\dagger = H$, because $(D^{+})^\dagger = -D^{-}$ holds.
By performing the Hamiltonian simulation algorithm for this Hamiltonian, we can simulate the one-dimensional wave equation on a quantum computer.

\subsection{Time evolution operator}
To derive a quantum circuit for time evolution by the Hamiltonian $H$, we first expand the Hamiltonian as follows:
\begin{align}
    H &= c \left( \sigma_{01} \otimes D^{+} - \sigma_{10} \otimes D^{-} \right) \nonumber \\
    &= \frac{c}{l} \sum_{j=1}^{n} \left( \sigma_{01} \otimes I^{\otimes (n-j)} \otimes \sigma_{01} \otimes \sigma_{10}^{\otimes (j-1)} \right. \nonumber \\
    &\quad \quad \quad \quad \left. + \sigma_{10} \otimes I^{\otimes (n-j)} \otimes \sigma_{10} \otimes \sigma_{01}^{\otimes (j-1)} \right) \nonumber \\
    &\quad - \frac{c}{l} (\sigma_{01} + \sigma_{10}) \otimes I^{\otimes n} \nonumber \\
    &=\frac{c}{l} \sum_{j=0}^{n} h_j,
\end{align}
where $h_{j>0} := \sigma_{01} \otimes I^{\otimes (n-j)} \otimes \sigma_{01} \otimes \sigma_{10}^{\otimes (j-1)} + \sigma_{10} \otimes I^{\otimes (n-j)} \otimes \sigma_{10} \otimes \sigma_{01}^{\otimes (j-1)}$ and $h_0 := -(\sigma_{01} + \sigma_{10}) \otimes I^{\otimes n}$.
The operator norm of the commutator of $h_j$ and $h_{j'}$ reads
\begin{align}
    \| [h_j, h_{j'}] \| &= 0 && \text{ for } j, j' \geq 1, \nonumber \\
    \| [h_j, h_0] \| &= 1 && \text{ for } j \geq 1.
\end{align} 
This means that the terms of the Hamiltonian $H$ can be grouped into two sets based on the commutator relation, in which one term ($h_0$) does not commute with the others, while the others commute with each other.   %
Let us define $H_1= (c/l) h_0$ and $H_2 = \sum_{j=1}^n (c/l) h_j$.
The time evolution operator with a time increment $\tau$ will then end up with a two-term product formula by approximating it using the first-order Trotter approximation, as follows:
\begin{align}
    \exp (-i H \tau) = \exp (-i H_1 \tau) \exp (-i H_2 \tau) + \mathcal{O}(\tau^2).
    \label{eq:exp_H}
\end{align}
This approximation has the accuracy of the second-order formula, because the Hamiltonian consists of two non-commutative terms~\cite{layden2022first}.  Using $\exp (-i H_1 \tau) \exp (-i H_2 \tau)$ as a one-step time evolution operator, we can obtain the evolved state at time $T$ by repeatedly applying the operator $T/\tau$ times.

\subsection{Quantum circuits for time evolution}
\label{sec:circ_teop}

\begin{figure}[htb]
\centering
\includegraphics[width=\columnwidth]{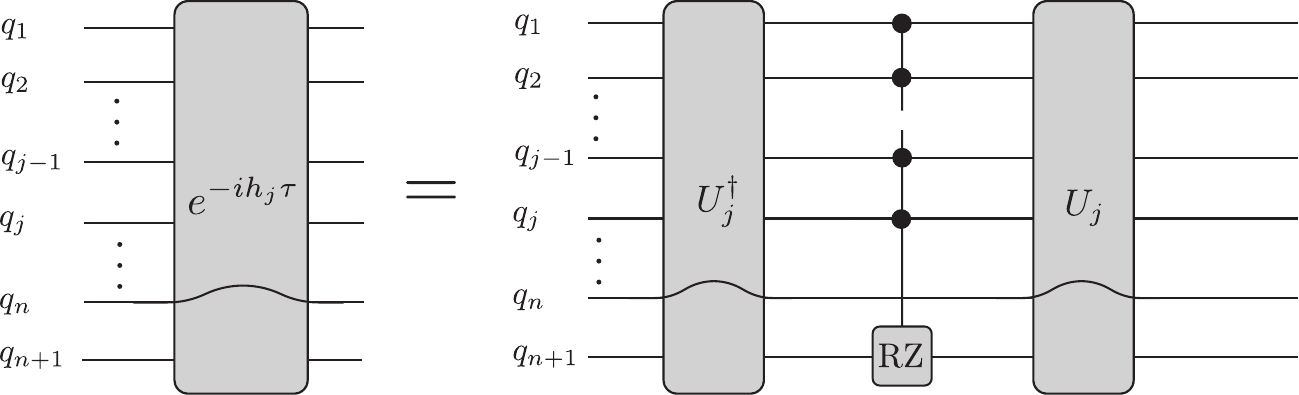}
\caption{Quantum circuit for $\exp (-i cl^{-1} h_j \tau)$, which operates on the $1,\ldots,j$- and $(n+1)$-th qubits. 
The rotation angle for the multi-controlled RZ gate is $2cl^{-1}$.}
\label{fig:circuit_hsim1}
\end{figure}

We now build a concrete quantum circuit for the one-step time evolution operator. First, we can implement $\exp (-i H_1 \tau)$ simply as $\texttt{RX} (-2cl^{-1} \tau) \otimes I^{\otimes n}$ where $\texttt{RX}(\theta)$ is a $\theta$ radian rotation about the X axis. 
Next, to derive a circuit for $\exp (-i H_2 \tau)$, we first diagonalize the $j$-th term of $H_2$, with omitting the identity operators, as follows.
\begin{align}
h_j    &= \ket{0}_{n+1} \ket{0}_j \! \ket{1}^{\otimes (j-1)} \! \bra{1}_{n+1} \bra{1}_j \! \bra{0}^{\otimes (j-1)} \nonumber \\
    &\quad + \ket{1}_{n+1} \ket{1}_j \! \ket{0}^{\otimes (j-1)} \! \bra{0}_{n+1} \bra{0}_j \! \bra{1}^{\otimes (j-1)} \nonumber \\
    &= \frac{ \ket{0}_{n+1} \ket{0}_j \! \ket{1}^{\otimes (j-1)} + \ket{1}_{n+1} \ket{1}_j \! \ket{0}^{\otimes (j-1)} }{\sqrt{2}} \nonumber \\
    &\quad \quad \times \frac{ \bra{0}_{n+1} \bra{0}_j \! \bra{1}^{\otimes (j-1)} + \bra{1}_{n+1} \bra{1}_j \! \bra{0}^{\otimes (j-1)} }{\sqrt{2}} \nonumber \\
    &\quad - \frac{ \ket{0}_{n+1} \ket{0}_j \! \ket{1}^{\otimes (j-1)} - \ket{1}_{n+1} \ket{1}_j \! \ket{0}^{\otimes (j-1)} }{\sqrt{2}} \nonumber \\
    &\quad \quad \times \frac{ \bra{0}_{n+1} \bra{0}_j \! \bra{1}^{\otimes (j-1)} - \bra{1}_{n+1} \bra{1}_j \! \bra{0}^{\otimes (j-1)} }{\sqrt{2}} \nonumber \\
    &=U_j \left( Z_{n+1} \otimes \ket{1} \! \bra{1}^{\otimes j} \right) U_j^\dagger,
\label{eq:h_j}
\end{align}
where $Z_{n+1}$ is the Pauli Z operator acting on the $(n+1)$-th qubit and $U_j$ is a unitary operator that transforms two states as follows.
\begin{align*}
   U_j \ket{0}_{n+1} \ket{1}^{\otimes j} &= \frac{ \ket{0}_{n+1} \ket{0}_j \! \ket{1}^{\otimes (j-1)} + \ket{1}_{n+1} \ket{1}_j \! \ket{0}^{\otimes (j-1)} }{\sqrt{2}}, \nonumber \\
   U_j \ket{1}_{n+1} \ket{1}^{\otimes j} &= \frac{ \ket{0}_{n+1} \ket{0}_j \! \ket{1}^{\otimes (j-1)} - \ket{1}_{n+1} \ket{1}_j \! \ket{0}^{\otimes (j-1)} }{\sqrt{2}}. \nonumber \nonumber \\ 
\end{align*}
Note that the circuit for $U_j$ can be built similarly to the one for preparing the GHZ state. 
With Eq.~(\ref{eq:h_j}), we can make the following decomposition,
\begin{align*}
 \exp (-i cl^{-1} h_j \tau) =  U_j \texttt{CRZ}_{n+1}^{1, \dots, j} (2cl^{-1}) U_j^\dagger, \label{eq:exp_hj}
\end{align*}
where $\texttt{CRZ}_{n+1}^{\bm{b}}(\theta)$ is a $\theta$ radian rotation gate about the Z axis on the $(n+1)$-th qubit controlled by all the qubits in $\bm{b}$, leading to the circuit shown in 
Figure~\ref{fig:circuit_hsim1}. Finally, by concatenating these circuits for $1 \le j \le n$, we have a circuit for $\exp (-i H_2 \tau)$.

\begin{figure}[htb]
\centering
\includegraphics[width=\columnwidth]{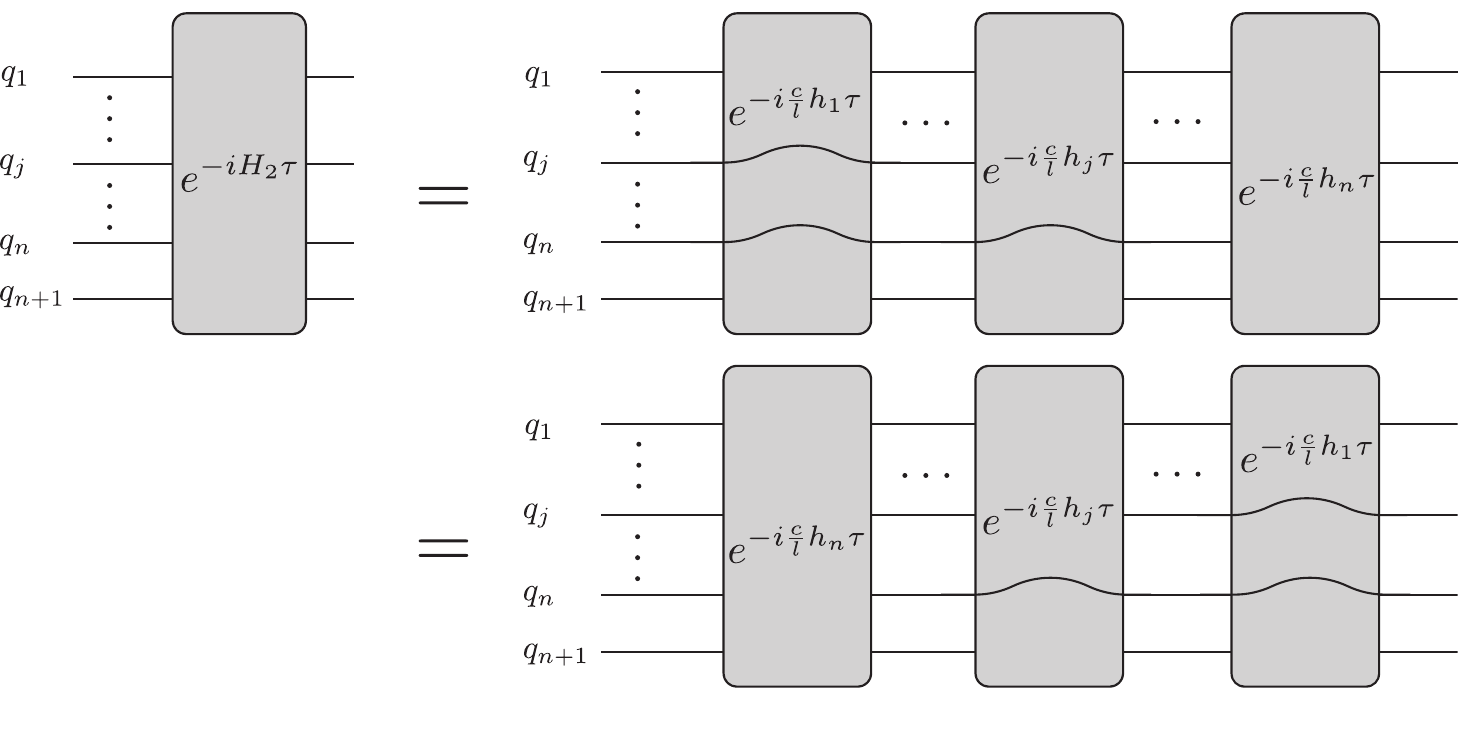}
\caption{Two simplest quantum circuits for $\exp (-i H_2 \tau)$. The circuits of $\exp (-i cl^{-1} h_j \tau) (1 \le j \le n)$ are concatenated in the increasing order (above) 
and the decreasing order (below).
}
\label{fig:circuit_hsim2}
\end{figure}

\begin{figure}[htbp]
  \centering
  \begin{subfigure}[b]{0.23\textwidth}
    \centering
    \includegraphics[width=\textwidth]{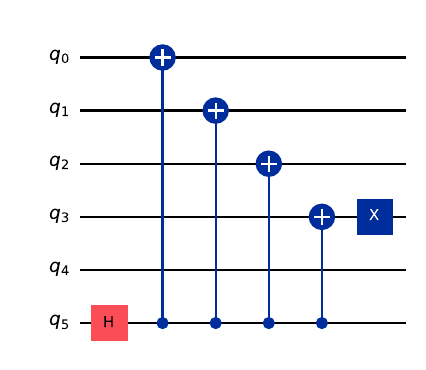}
    \caption{spray-typed}
  \end{subfigure}
  \hfill
  \begin{subfigure}[b]{0.23\textwidth}
    \centering
    \includegraphics[width=\textwidth]{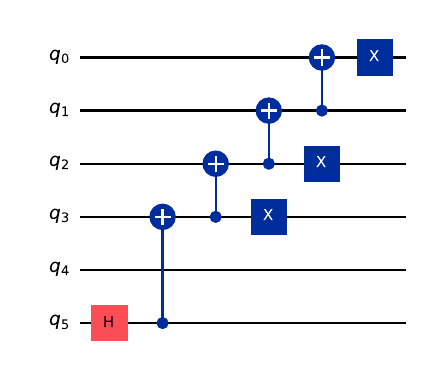}
    \caption{stair-typed}
  \end{subfigure}
  
  \caption{Two circuits to implement $U_4$ for the 6-qubit circuit, which we call spray-typed (left) and stair-typed (right) based on the shapes formed by the CX gates. The Hadamard and X gates are in red and navy blue, respectively. }
  \label{fig:U4}
\end{figure}

It should be noted that numerous different circuits can implement $\exp (-i H \tau)$ in Eq.~(\ref{eq:exp_H}). 
For instance, the concatenation of the $n$ circuits 
of $\exp (-i cl^{-1} h_j \tau) (1 \le j \le n)$
for $\exp (-i H_2 \tau)$ can be of the arbitrary order.  We show two simplest ones in Figure~\ref{fig:circuit_hsim2}, which we call the increasing $H_2$ and the decreasing $H_2$.
In addition,
there are many different circuits to implement $U_j$, since only what we require is that $U_j$ transforms the two states as specified above.
We show two circuits to implement $U_4$ for the 6-qubit circuit (i.e., 5 qubits for discretization)
in Figure~\ref{fig:U4}, which we call spray-typed and stair-typed.  We will soon see that, while all these circuits perform exactly the same (approximated) time evolution, the circuit optimization ends up with generating very different circuits.

\section{Multilevel Circuit Optimization}
\label{sec:mlco}

{\it Multi-level circuit optimization} (MLCO) transforms the {\it source circuit} through multiple levels of gate sets for optimization, from higher to lower, to the {\it target circuit}.  At each level, we first perform circuit simplification, and then lower the simplified circuit into the next level.  The source circuit is represented in the highest level, while the target circuit is represented in the lowest level.

We explore MLCO, assuming that some simplifications and optimizations are best performed at a higher level, while the opportunities for them are lost or obscure at a lower level. This follows the tradition of the (classical) compiler community where they deploy multilevel IRs (Intermediate Representations) to allow optimizations to be implemented at the IR levels best suited for them~\cite{stanier2013IRs}.
In contrast to MLCO, the current circuit compilers tend to decompose the source circuit into single- and two-qubit gates earlier, and direct their optimization efforts to the low-level circuit, expecting the resulting circuit to be highly optimized. We call this the DETO (DEcompose, Then Optimized) approach. We will compare the two in a quantitative manner in Section~\ref{sec:results}.

For our case study, we deploy three gate sets: high-, mid-, and low-level. They are different in how many control qubits are allowed in their gates.\footnote{We consider a single-qubit gate to be a gate controlled by {\it zero} qubit.}
The high-level gate set (HiGS) allows the arbitrary number of control qubits. It is used to represent the source circuit to solve a PDE without decomposing any multi-controlled gates.  The mid-level gate set (MiGS) allows two or fewer control qubits.  In particular, it includes the Toffoli gate, which has been long studied and proven to be highly expressive and amenable to simplification. Finally, the low-level gate set (LoGS) only allows one or zero control qubits, similar to the hardware gate sets of most devices available today. It is used to represent the target circuit. Note that we do not care about what specific gates the gate set contains, as long as it is universal and satisfies the condition on the number of controls.

As we will soon see, every controlled gate has a single-qubit target in our case study. 
When the source circuit contains multi-target multi-controlled gates, we may also have to define different gate sets based on the number of target qubits. Defining such gate sets based on case studies will constitute an interesting direction for future research.

In this paper, we do not further compile the target circuit into the hardware gate set of a particular device. In other words, we do not deal with qubit mapping and routing. Our focus is to optimize circuits to minimize the number of two-qubit gates in the target circuit. 

Our case study is on quantum circuits for Hamiltonian simulation to solve a one-dimensional wave equation which we explained in the previous section. 
To make the discussion concrete and clear,
we consider the 6-qubit circuits (5 qubits for discretization) with $\tau=0.2$, $l=1$, and $c=1$.

\subsection{Building Circuits for One Trotter Step}

\begin{figure*}[ptb]
  \centering
  \begin{subfigure}[b]{\textwidth}
    \centering
    \includegraphics[width=\textwidth]{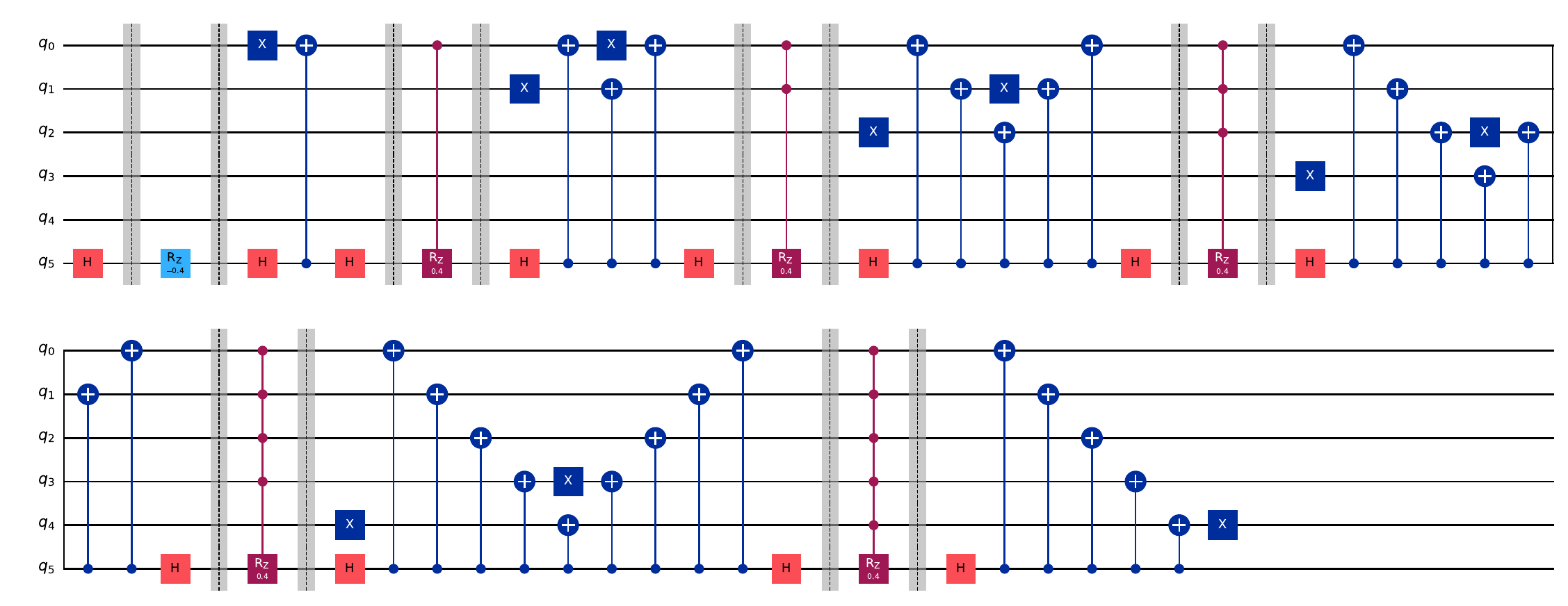}
    \caption{Source circuit with spray-typed wings}
    \label{fig:higs_spray}
  \end{subfigure}
  \hfill
  \begin{subfigure}[b]{\textwidth}
    \centering
    \includegraphics[width=\textwidth]{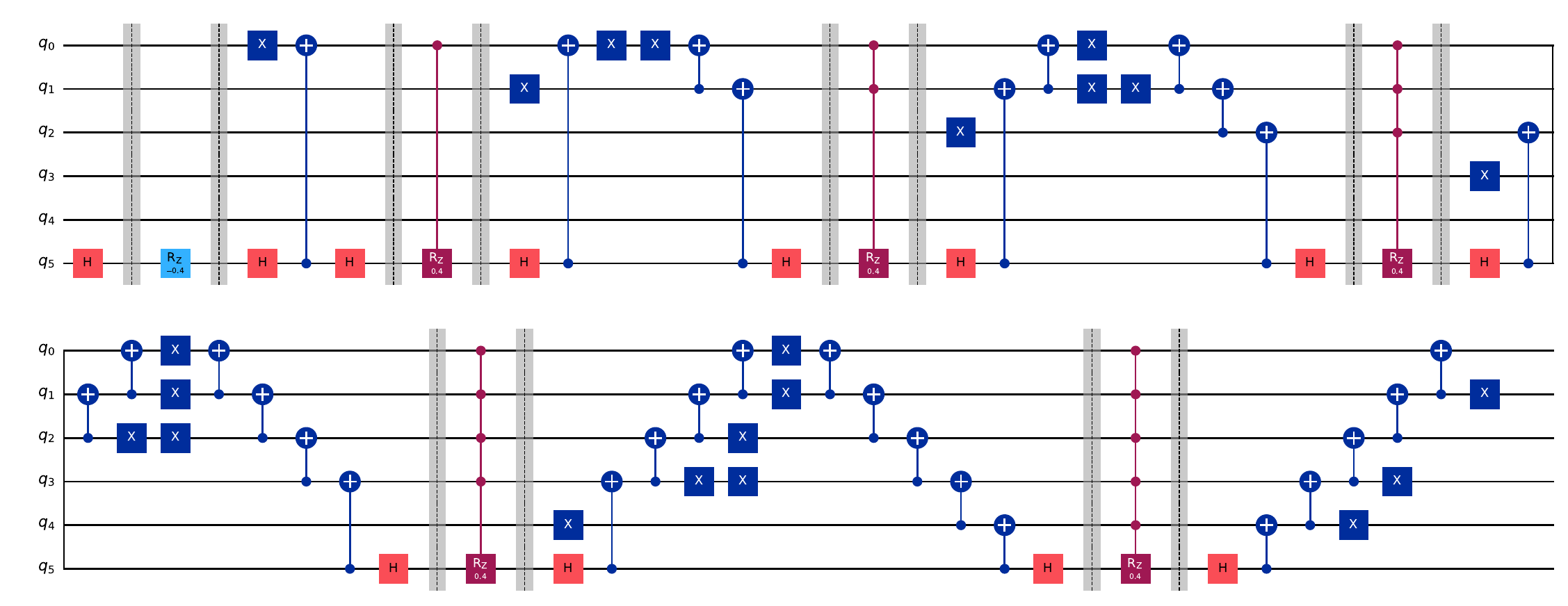}
    \caption{Source circuit with stair-typed wings}
    \label{fig:higs_stair}
  \end{subfigure}
  
  \caption{Two source circuits of 6 qubits (5 qubits for discretization) for one-step time evolution. The wings of multi-controlled RZ gates are (a) spray-typed and (b) stair-typed. In both, the controlled RZ gates are in reddish brown, while the RZ gate near the beginning is in blue. The Hadamard and X gates are in red and navy blue, respectively.  The first five single-qubit gates at Qubit 5 are from $H_1$. The rest of the gates are from $H_2$, where the gates to represent $\exp (-i cl^{-1} h_j \tau)$ are placed in the increasing order of $j~(1 \le j \le 5)$.  Note that barriers are inserted around the backbones to improve the visibility of the circuits.}
  \label{fig:source}
\end{figure*}

As we discussed in Section~\ref{sec:circ_teop}, we have many possible source circuits for our Hamiltonian simulation.  We study two of them in Figure \ref{fig:source}, one with spray-typed for $U_j$
and the other with stair-typed. 
Both adopt the increasing $H_2$.  Note that barriers are placed to improve the visibility of the circuits.
We call each multi-controlled RZ gate a backbone, and the corresponding sub-circuits for $U_j^\dagger$ and $U_j$ its left and right wings, respectively.

\subsection{Simplifying Circuits at the HiGS}

\begin{figure*}%
  \centering
  \begin{subfigure}[b]{\textwidth}
    \centering
    \includegraphics[width=\textwidth]{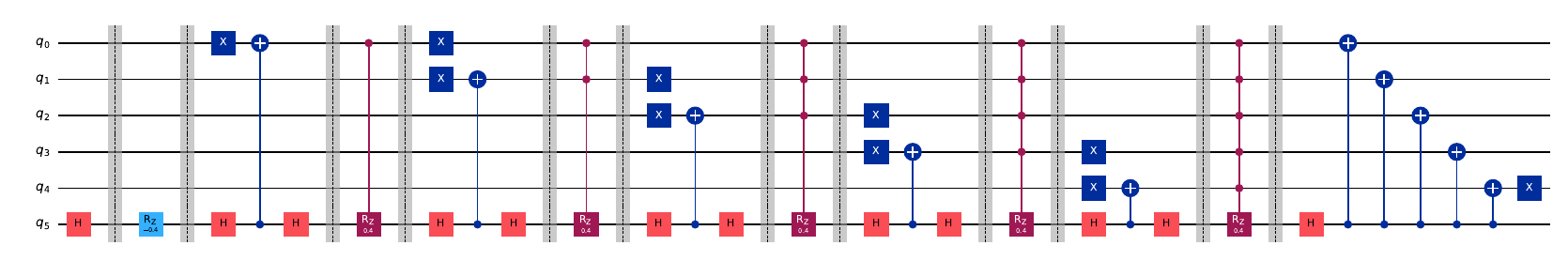}
    \caption{Simplified circuit at HiGS with spray-typed wings}
    \label{fig:hsim_spray}
  \end{subfigure}
  \hfill
  \begin{subfigure}[b]{\textwidth}
    \centering
    \includegraphics[width=\textwidth]{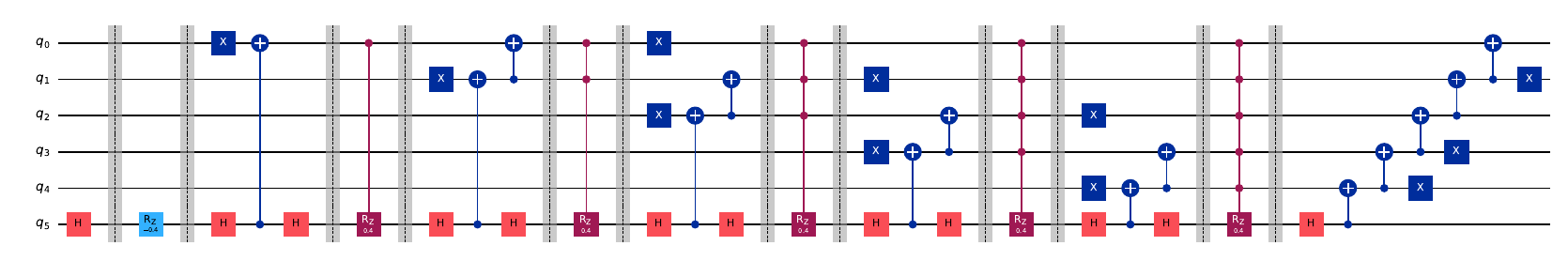}
    \caption{Simplified circuit at HiGS with stair-typed wings}
    \label{fig:hsim_stair}
  \end{subfigure}
  
  \caption{The resulting circuits after simplifying the source circuits primarily through gate commutation and cancellation.  Massive cancellation happened between the CX gates in the right wing of one backbone and those in the left wing of the next backbone. We thus only see a remnant of wings. Note that barriers are inserted around the backbones to improve the visibility of the circuits.}
  \label{fig:hsim}
\end{figure*}

\begin{figure}%
  \centering
  \begin{subfigure}[b]{0.20\textwidth}
    \centering
    \includegraphics[width=0.95\linewidth]{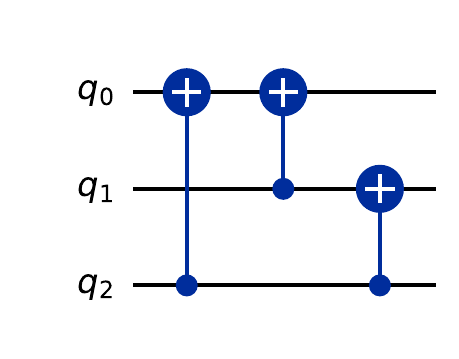}
  \end{subfigure}
  \begin{subfigure}[b]{0.158\textwidth}
    \centering
    \includegraphics[width=0.95\linewidth]{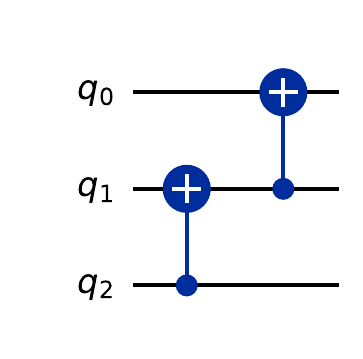}
  \end{subfigure}
  
  \caption{Two equivalent circuits. The left circuit with three CX gates can be reduced with the right circuit with two CX gates.}
  \label{fig:cx3cx2}
\end{figure}

For each circuit, the first task at the HiGS is to simplify it.  Figure \ref{fig:hsim} shows the resulting circuits. The simplification is performed through gate commutation and cancellation, and additionally the equivalence relation in Figure \ref{fig:cx3cx2} for the stair-typed.  In both circuits, massive cancellation occurred between the CX gates in the right wing of one backbone and those in the left wing of the next backbone. 

The second task at the HiGS is to build the output circuit represented in the MiGS, that is, to decompose multi-controlled gates with more than two control qubits into the MiGS. We describe this in detail in the next subsection.

\subsection{Decomposing Multi-Controlled Gates}

Decomposing multi-controlled (single-target) gates has been an extensive area of research since the early days of quantum computing~\cite{barenco1995elementary}.
Numerous algorithms have been proposed, some of which use auxiliary qubits like this study, e.g.~\cite{selinger2013quantum,jones2013low,amy2014polynomial,maslov2016advantages,he2017decompositions,rosa2025optimizing}, while others do not, e.g.~\cite{saeedi2013linear,iten2016quantum,vale2023circuit}.
Note that not only does the use of auxiliary bits vary, but also what is optimized, whether it is the circuit depth or the number of gates, differs among algorithms.
As a result, the modern circuit compiler supports multiple decomposition algorithms, although not necessarily many.

\begin{figure}%
  \centering
  \begin{subfigure}[t]{0.20\textwidth}
    \centering
    \includegraphics[width=0.50\linewidth]{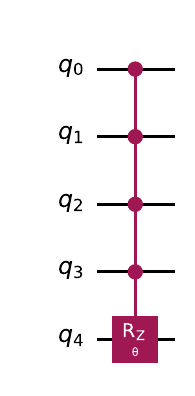}
  \end{subfigure}
  \begin{subfigure}[b]{0.20\textwidth}
    \centering   
    \includegraphics[width=0.95\linewidth]{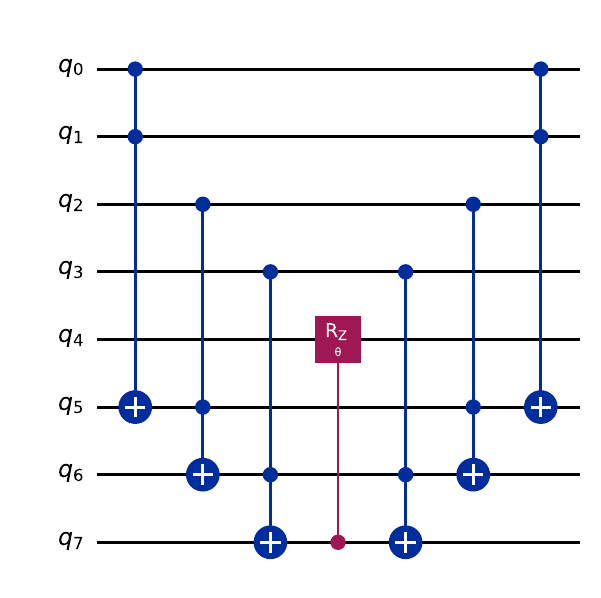}
  \end{subfigure}
  
  \caption{The Z rotation gate with four control qubits (left) and its {\it vchain} decomposition (right).}
  \label{fig:vchain}
\end{figure}

One of the simplest algorithms to decompose a multi-controlled gate, introduced in many textbooks and sometimes called {\it vchain}, uses a number of Toffoli gates and a number of clean auxiliary qubits, both proportional to the number of control qubits~\cite{nielsen2010quantum}. For example, Figure~\ref{fig:vchain} shows how it decomposes the Z rotation gate controlled by four qubits ($q_0$--$q_4$), using three clean auxiliary qubits ($q_5$--$q_7$). The Toffoli gates are symmetrically placed in both {\it flanks} of the controlled Z rotation gate.

\begin{figure*}%
  \centering
  \begin{subfigure}[b]{\textwidth}
    \centering
    \includegraphics[width=\textwidth]{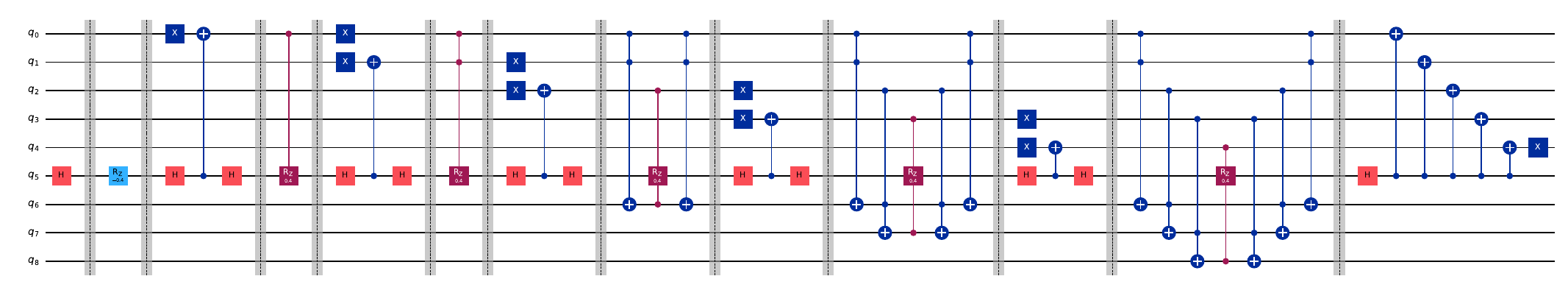}
    \caption{Input circuit at MIGS with spray-typed wings}
    \label{fig:migs_spray}
  \end{subfigure}
  \hfill
  \begin{subfigure}[b]{\textwidth}
    \centering
    \includegraphics[width=\textwidth]{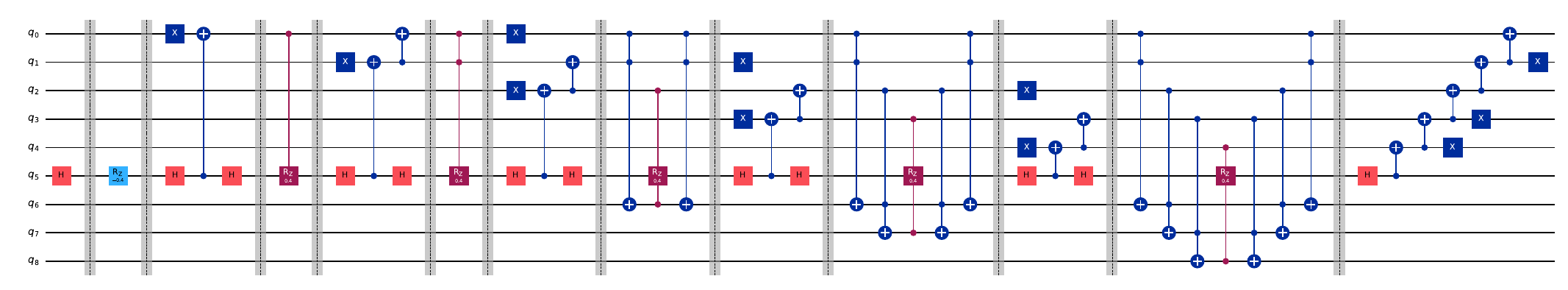}
    \caption{Input circuit at MiGS with stair-typed wings}
    \label{fig:migs_stair}
  \end{subfigure}
  
\caption{The resulting circuits after lowering HiGS circuits into MiGS, decomposing multi-controlled gates with more than two control qubits by a variant of {\it vchain}. Note that we keep the CCRZ gates, and that barriers are inserted around the backbones, whether intact or decomposed, to improve the visibility of the circuits.}
  \label{fig:migs}
\end{figure*}

While many algorithms are known, it is up to the programmer to pick which one to use for their circuit. We claim that the decision should be made on the basis of the whole circuit structure.  For our PDE circuits, we chose a variant of the {\it vchain} decomposition, lowering the circuits in Figure~\ref{fig:hsim_spray} and \ref{fig:hsim_stair} into those shown in Figure~\ref{fig:migs_spray} and \ref{fig:migs_stair}, respectively. Note that we keep the CCRZ gates (the Z rotation gates controlled by two qubits) in our variant. 

As clearly observed in these circuits, a significant amount of opportunities for circuit simplification has been created with regard to the Toffoli gates.  This is the reason why we chose the {\it vchain}, together with the assumption that there will be no shortage of qubits, which we believe is reasonable for many applications on the current noisy devices. 

\subsection{Simplifying Circuits at the MiGS}

\begin{figure*}%
  \centering
  \begin{subfigure}[b]{\textwidth}
    \centering
    \includegraphics[width=0.70\textwidth]{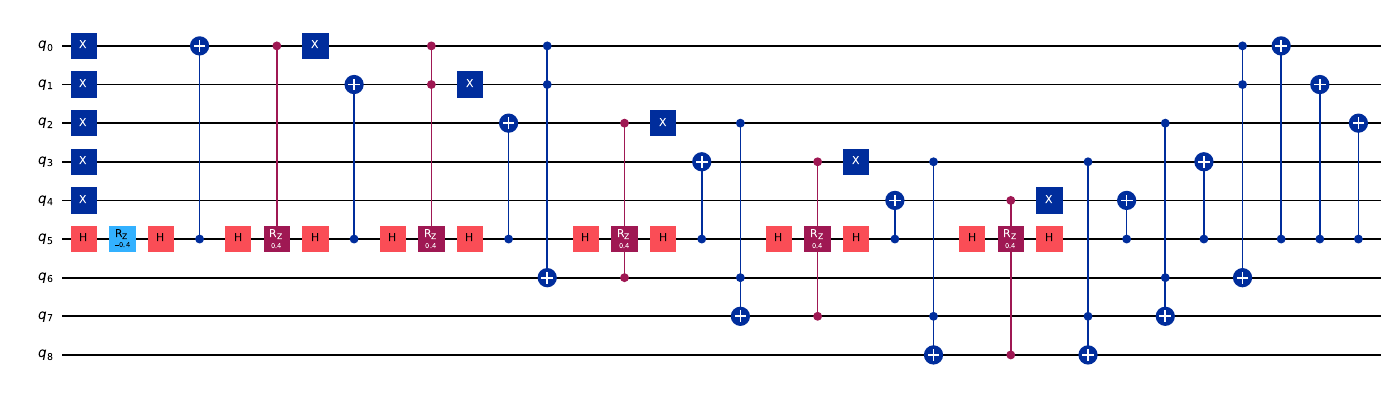}
    \caption{Simplified MiGS circuit with spray-typed wings}
    \label{fig:msim_spray}
  \end{subfigure}
  \hfill
  \begin{subfigure}[b]{\textwidth}
    \centering
    \includegraphics[width=0.70\textwidth]{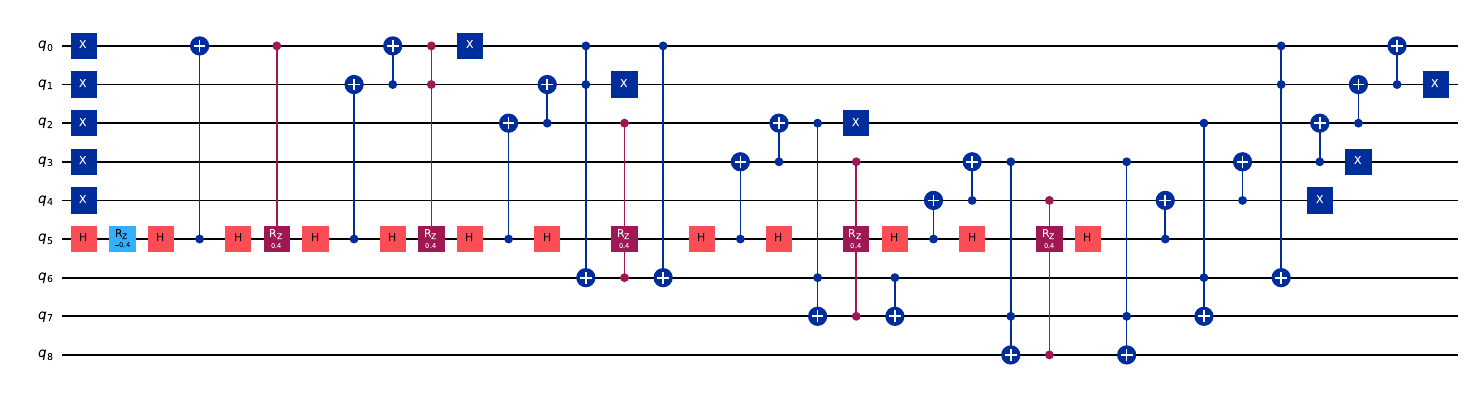}
    \caption{Simplified circuit at MiGS with stair-typed wings}
    \label{fig:msim_stair}
  \end{subfigure}
  
  \caption{The resulting circuits after simplifying the input circuits at the MiGS primarily through gate commutation and cancellation.  Massive cancellation happened between the Toffoli gates in the right flank of one {\it vchain} decomposition and those in the left flank of the next. However, an intact block of entangling gates still remains at the end of the circuits. Note that barriers are no longer inserted, resulting in the Qiskit transpiler having reordered gates.}
  \label{fig:msim}
\end{figure*}

\begin{figure*}%
  \centering
  \begin{subfigure}[b]{\textwidth}
    \centering
    \includegraphics[width=0.70\textwidth]{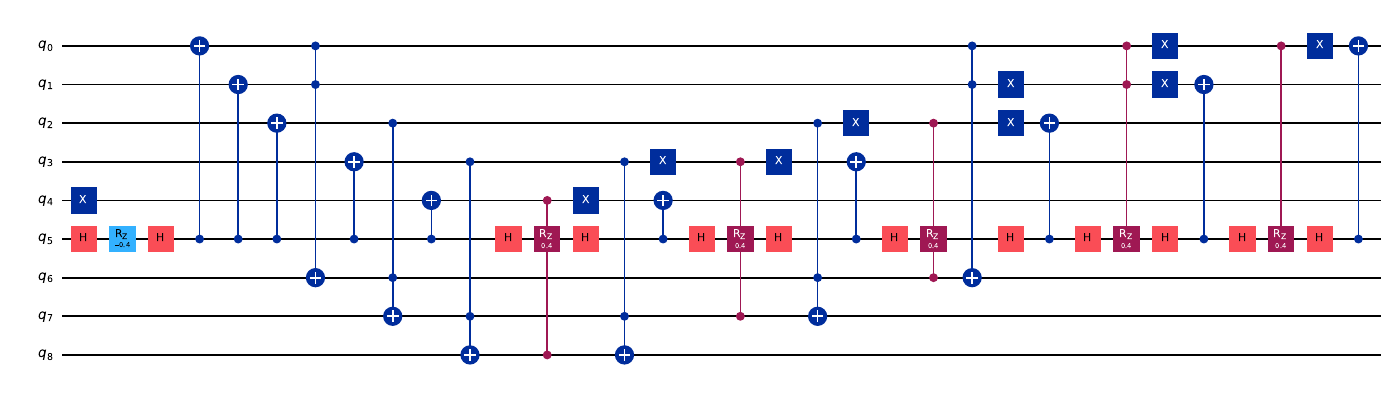}
    \caption{Simpfilied circuit at MiGS with the decreasing $H_2$ (spray-typed)}
    \label{fig:msim_dH2_spray}
  \end{subfigure}
  \hfill
  \begin{subfigure}[b]{\textwidth}
    \centering
    \includegraphics[width=0.70\textwidth]{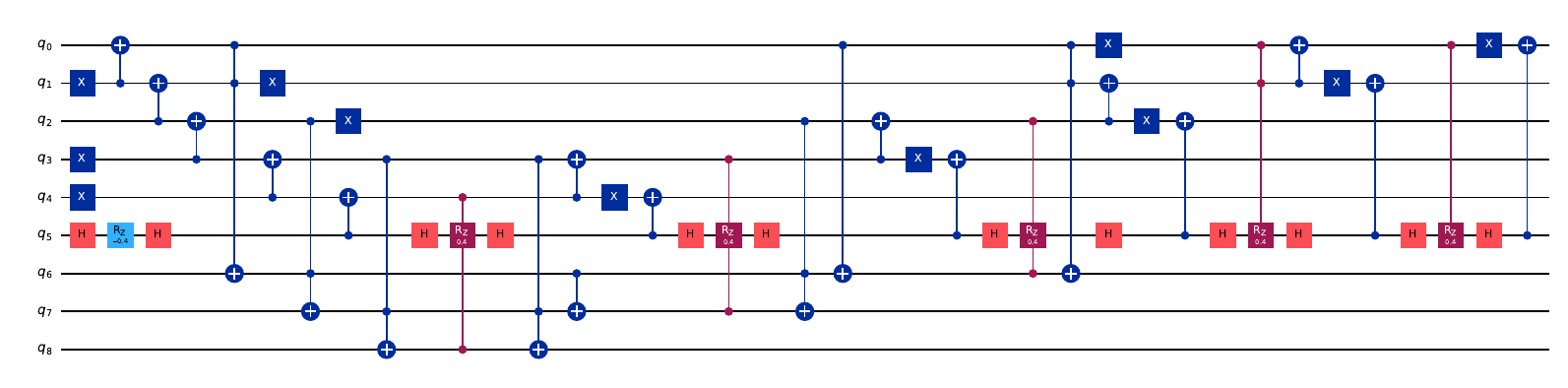}
    \caption{Simplified circuit at MiGS with the decreasing $H_2$ (stair-typed)}
    \label{fig:msim_dH2_stair}
  \end{subfigure}
  
\caption{The circuits built with the decreasing $H_2$ after simplification at the MiGS.
An intact block of entangling gates now appears near the beginning of each circuit.}
  \label{fig:msim_dH2}
\end{figure*}
\begin{figure}[htbp]
  \centering
  \begin{subfigure}[t]{0.20\textwidth}
    \centering
    \includegraphics[width=0.95\linewidth]{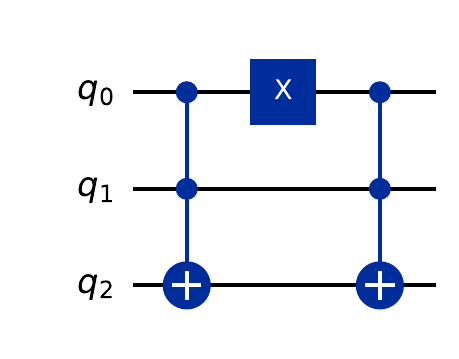}
  \end{subfigure}
  \begin{subfigure}[b]{0.20\textwidth}
    \centering   
    \includegraphics[width=0.55\linewidth]{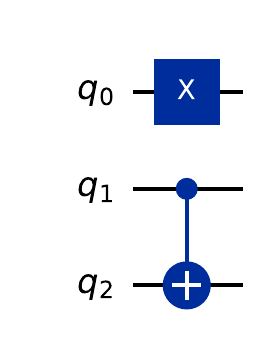}
  \end{subfigure}
  
\caption{Two equivalent circuits. The left circuit with two CCX gates can be significantly reduced to the right circuit with just one CX gate.}
  \label{fig:ccx2cx}
\end{figure}

Performing simplification at the MiGS, we have the circuits shown in Figure~\ref{fig:msim_spray} and \ref{fig:msim_stair}.  Please note that we used the equivalence relation in Figure~\ref{fig:ccx2cx} to simplify the stair-typed circuit.  In both circuits, massive cancellation occurred between the Toffoli gates in the right flank of one {\it vchain} decomposition and those in the left flank of the next.

We now explain why we keep the CCRZ gates.  Assume that we replace the CCRZ gate with the sequence of CCX (Toffoli), CZ, and CCX gates as in Figure~\ref{fig:vchain}. We now understand that these CCX gates are not subject to the cancellation we have just observed.  Since the CCRZ gate is decomposed into the gate sequence with four CX gates, the above result is not as efficient.

As you see, the spray-typed circuit is simpler than the stair-typed one for one-step time evolution. In either case, the output circuits have been significantly simplified, compared to the source circuits in the HiGS. However, there remains an intact block of entangling gates at the end of each circuit, which are from both the right wing and frank of the last controlled RZ gate. The intact block is our next target for circuit optimization.

\subsection{Building Circuits for Two Trotter Steps}

To simulate time evolution by a Hamiltonian via Trotter approximation, it is common that we first create a circuit to prepare the initial state and then append to it a circuit for one step as many times as necessary.  For our case study, observing that we have many different circuits for one step (two of which are shown above), we can mix them to build a circuit for longer time steps.  While these may not be equivalent to each other, they exhibit the same behavior for evolution of one time step.

Along with this line of thought, we build a circuit for two Trotter steps by composing a one-step circuit with the increasing $H_2$ and a one-step circuit with the decreasing $H_2$.  We can do this both for the spray- and stair-typed. 

To begin with, we show two types of one-step circuits with the decreasing $H_2$ in Figure~\ref{fig:msim_dH2}.  These have been simplified at the MiGS, similar to those in Figure~\ref{fig:msim}. As you see, there now remains an intact block of entangling gates near the beginning of each circuit, which are from the left wing and flank of the first controlled RZ gate.  

\begin{figure*}%
  \centering
  \begin{subfigure}[b]{\textwidth}
    \centering
    \includegraphics[width=\textwidth]{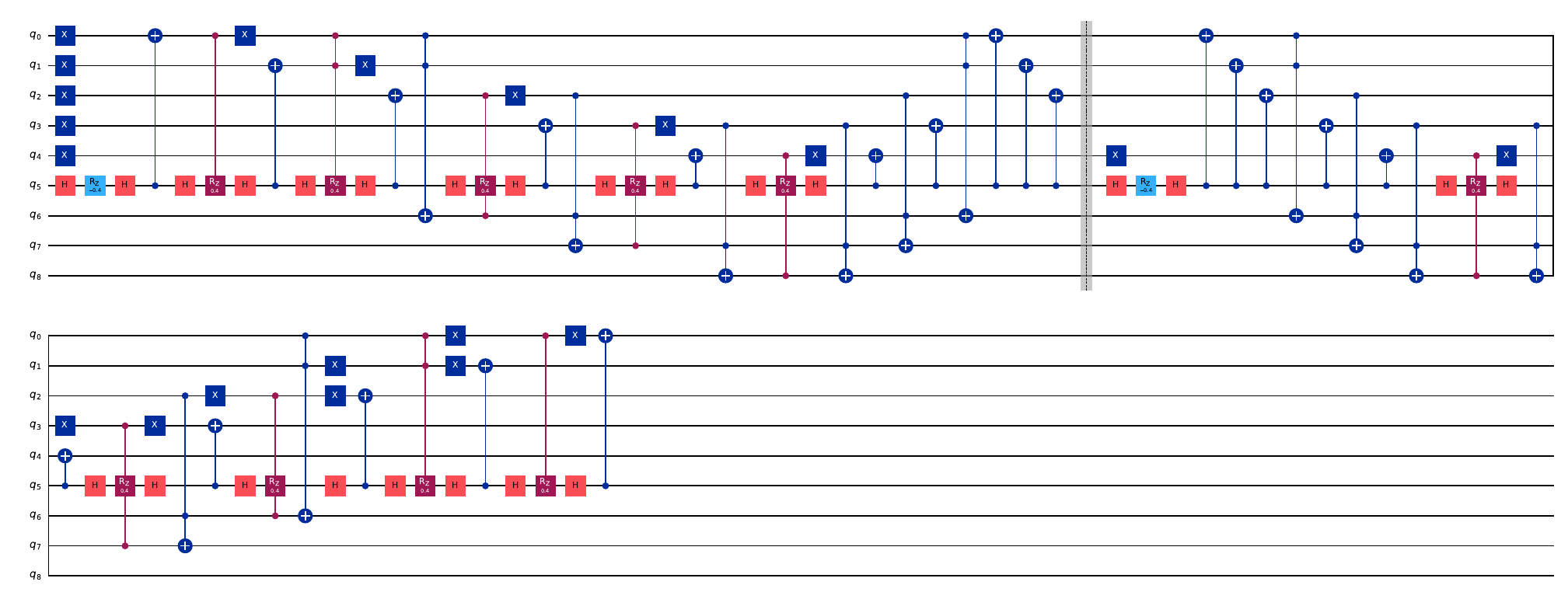}
    \caption{Two-step circuit at MIGS (spray-typed)}
    \label{fig:msim_spray2}
  \end{subfigure}
  \hfill
  \begin{subfigure}[b]{\textwidth}
    \centering
    \includegraphics[width=\textwidth]{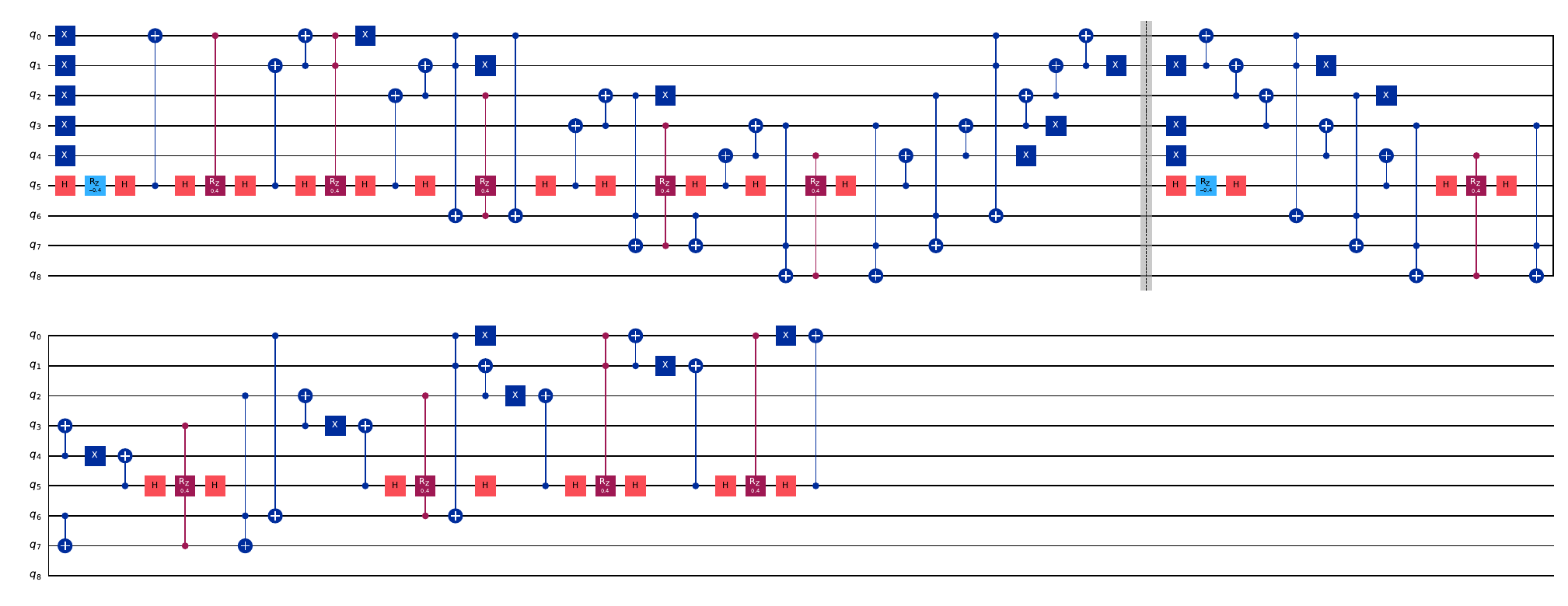}
    \caption{Two-step circuit at MiGS (stair-typed)}
    \label{fig:msim_stair2}
  \end{subfigure}
  
\caption{Two-step circuits represented in the MiGS, built 
by composing a one-step circuit with the increasing $H_2$ and a one-step circuit with the decreasing $H_2$.
The intact block of entangling gates of the former one-step circuit now faces that of the latter across the single-qubit gates from $H_1$.  Note that a barrier is inserted for better visibility between the two one-step circuits composed.}
\label{fig:msim_2step}
\end{figure*}

Figure~\ref{fig:msim_2step} shows the two-step circuits built in the above-mentioned manner for the spray-typed and stair-typed wings.
Note that, while we performed the composition after simplification at the MiGS, we can do so before the simplification or even at the HiGS. In each of the spray- and stair-typed, the two intact blocks of entangling gates now face each other across the single-qubit gates at Qubit 5 from $H_1$, expected to cancel each other. 

\subsection{Simplifying Circuits for Two Trotter Steps at the MiGS}

Contrary to the expectation, no cancellation happens for the spray-typed.  The intervening gates from $H_1$ prevent the CX gates on both sides, which all use Qubit~5 as their controls, from canceling each other, which in turn prevents the CCX gates on both sides from canceling each other.

\begin{figure*}[thbp]
\centering
\includegraphics[width=\textwidth]{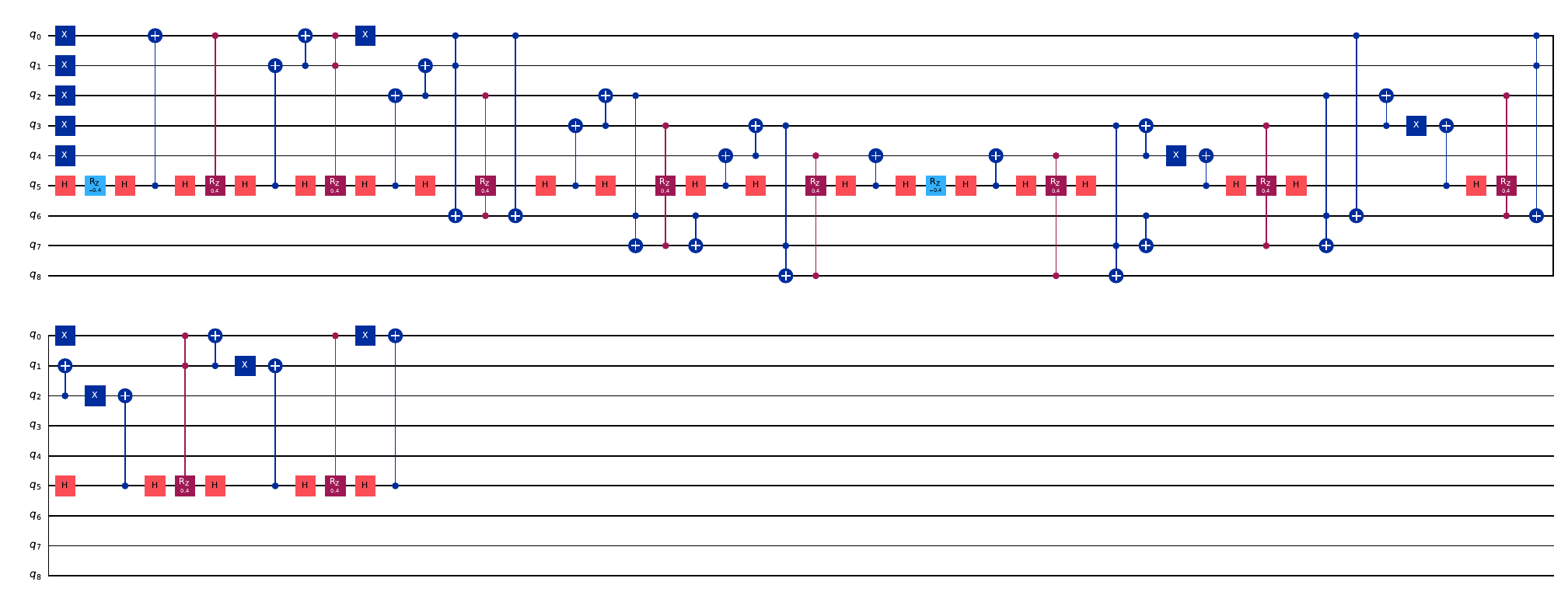}
\caption{Two-step circuit highly simplified at the MiGS. Massive cancellation happened for the formerly intact blocks of entangling gates. Only two CX gates survive which have dependence with the gates from $H_1$ on Qubit 5.}
\label{fig:msim2_stair2}
\end{figure*}

On the other hand, for the stair-typed, most of the CX gates of the intact blocks do not have dependence with Qubit~5, thereby canceling each other, which in turn enables the CCX gates on both sides to cancel each other. As a result, we have such a simplified circuit for the stair type as in Figure~\ref{fig:msim2_stair2}.

\subsection{Replacing Toffoli gates with relative-phase Toffoli gates}
\label{sec:ccx2rccx}

\begin{figure}[h]
 	\centering
	\includegraphics[width=\linewidth]{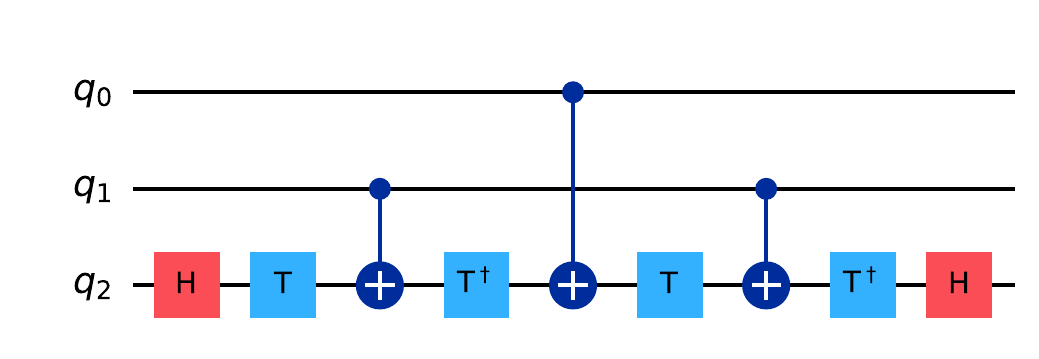}
	\caption{Decomposition of the relative-phase Toffoli (RCCX) gate. It is realized with three CX gates.}
\label{fig:rccx}
\end{figure}

An intriguing optimization specific at the MiGS is to explore replacement of the Toffoli gates with the relative-phase Toffoli gates (the RCCX gates)~\cite{barenco1995elementary,maslov2016advantages}. This is akin to the strength reduction of a classical compiler, since the Toffoli gate requires six CX gates on decomposition to single- and two-qubit gates, 
while the RCCX gate requires only three CX gates on such decomposition as in Figure~\ref{fig:rccx}.

\begin{figure}[htbp]
  \centering
  \begin{subfigure}[b]{0.23\textwidth}
    \centering
    \includegraphics[width=\linewidth]{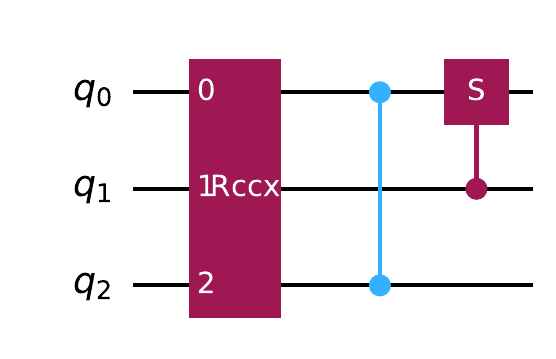}
  \end{subfigure}
  \begin{subfigure}[b]{0.23\textwidth}
    \centering
    \includegraphics[width=\linewidth]{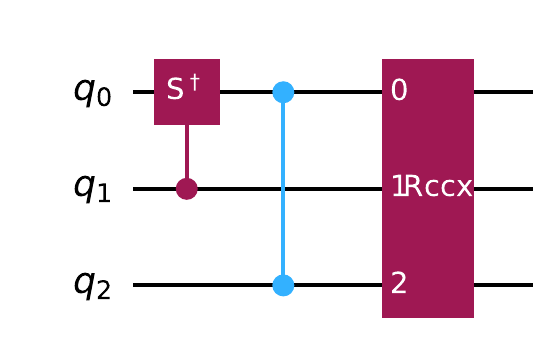}
  \end{subfigure}
  
  \caption{Two circuits equivalent to the Toffoli gate.}
  \label{fig:ccx2rccx}
\end{figure}

Concretely, consider a pair of Toffoli gates that act on the same control and target qubits. Observing that the Toffoli gate is equivalent to each of the two circuits shown in Figure~\ref{fig:ccx2rccx}, 
we attempt to replace the left and right Toffoli gates with the left and right circuits, respectively.
Then the CS (controlled S) and the C\Sdg gates cancel each other if the gates between the pair are diagonal or just read the qubits of the CS (and C\Sdg) gate.  If the two cancel each other, the same argument goes for the pair of CZ gates.

Let us turn our attention to the PDE circuit in Figure~\ref{fig:msim2_stair2}. We have three nested pairs of Toffoli gates.  Following the above arguments, only the RCCX gates remain for the innermost pair, while the RCCX and CZ gates stay for the outer pairs.

\begin{figure}[htbp]
  \centering
  \begin{subfigure}[b]{0.19\textwidth}
    \centering
    \includegraphics[width=0.95\linewidth]{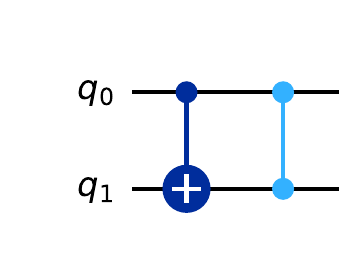}
  \end{subfigure}
  \begin{subfigure}[b]{0.19\textwidth}
    \centering
    \includegraphics[width=1.2\linewidth]{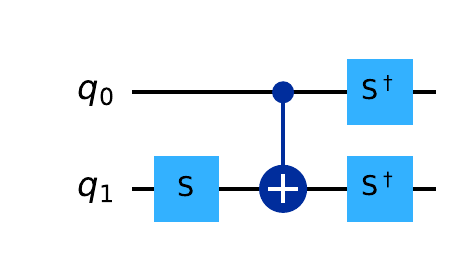}
  \end{subfigure}
  
  \caption{Two equivalent circuits. The left circuit with two entangling gates can be reduced to the right with one.}
  \label{fig:cxcz}
\end{figure}

A further optimization can be performed based on the fact that each of the remaining CZ gates finds the neighboring CX gate that acts on the same qubits. Then we can exploit the equivalence relations in Figure~\ref{fig:cxcz} for more reduction in the number of entangling gates. 

\begin{figure*}[thbp]
\centering
\includegraphics[width=\textwidth]{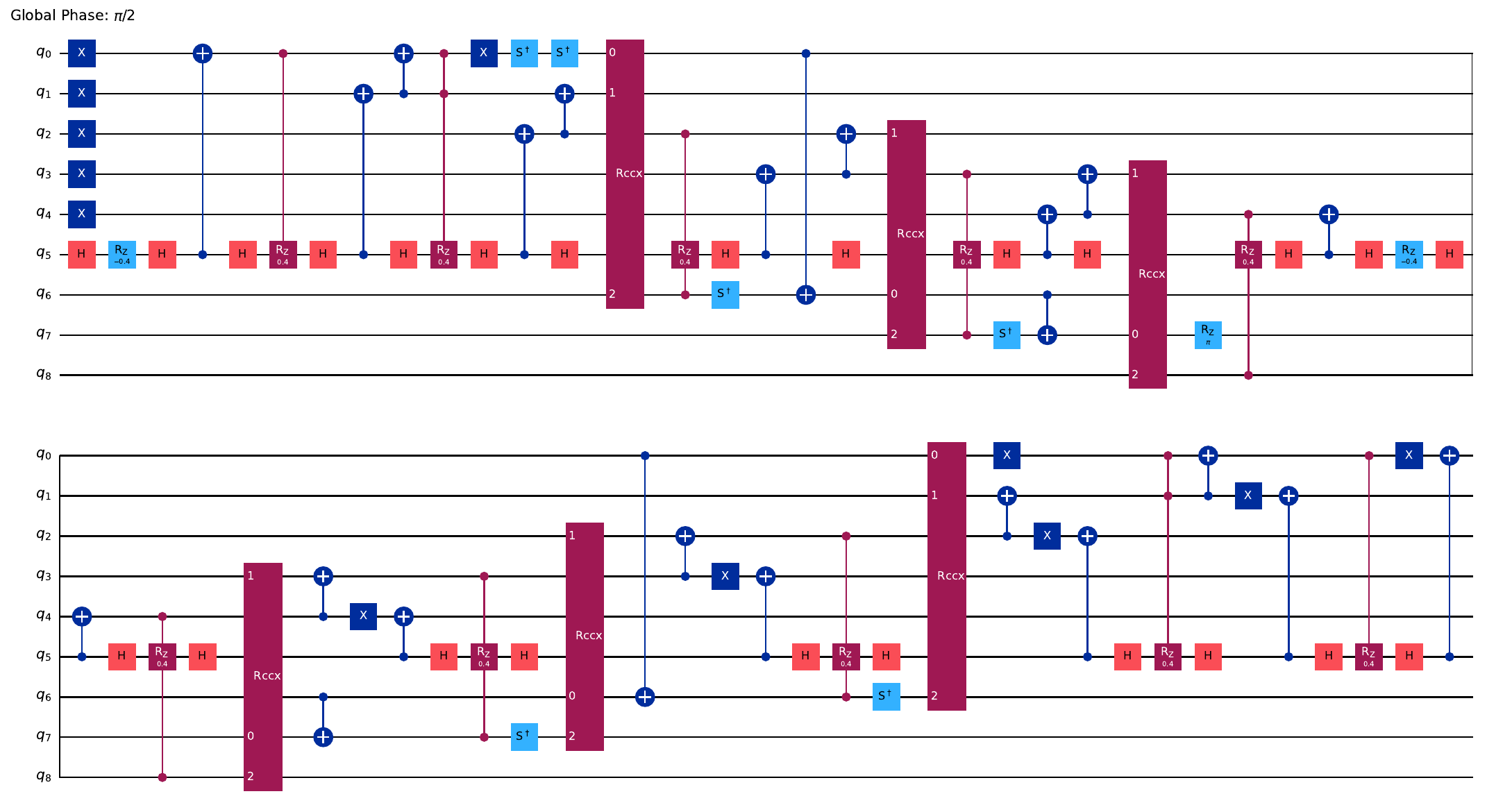}
\caption{Two-step circuit at the MiGS with the relative-phase Toffoli (RCCX) gates with which we replaced the Toffoli gates in Figure~\ref{fig:msim2_stair2}. }
\label{fig:msimRT2_stair2}
\end{figure*}

After all, we have the optimized two-step circuit at the MiGS as shown in Figure~\ref{fig:msimRT2_stair2}.

\subsection{Building and Optimizing Circuits at the LoGS}

Decomposing the CCX and RCCX gates in the MiGS circuit, we have the input circuit for the LoGS which consists of the single- and two-qubit gates.  Since the LoGS is similar to most of the hardware gate sets of devices available today, we can then rely on the existing modern circuit optimizer such as Qiskit~\cite{qiskit2024} for simplification and optimization at the LoGS.

\subsection{Structural Analysis}

We now discuss what structural characteristics
in our circuits caused such massive cancellation.
We think it is repeated occurrences of a pattern 
of $P^{\dagger} Q P$, where $P$ and $Q$ are blocks of gates in a circuit (the order is thus not that of matrix multiplication). We call this a {\it conjugate pattern}, two instances of which 
are found in Figure~\ref{fig:circuit_hsim1} (right) and Figure~\ref{fig:vchain} (right).
Two adjacent occurrences of a conjugate pattern are potential sources of cancellation, where the predecessor's $P$ and the successor's $P^{\dagger}$ may fully or partially cancel out each other.

There are two caveats. First, two adjacent occurrences do not necessarily act on the same set of qubits.  Thus, it is important to maximize the overlap as we do in the increasing $H_2$ and the decreasing $H_2$.
Second, many or a few gates intervene between the two adjacent occurrences, disrupting the cancellation.
Thus, it is important to eliminate as many of them as possible by simplification and optimization as we do in our case study.

\section{Implementation}
\label{sec:impl}

We implemented a prototype of our MLCO approach using Qiskit ~\cite{qiskit2024}, the most popular and powerful SDK for quantum computing. Specifically, we use Version 1.4.2 for our case study. %

We built the base circuit optimizer for each gate set by extending the predefined Pass Manager of the Qiskit transpiler.   For instance,  we created the base optimizer for the LoGS, which consists of the RZ, X, SX, and CX gates, as follows.

\begin{lstlisting}[language=Python, basicstyle=\ttfamily\small, frame=single]
from qiskit.transpiler import PassManager
from qiskit.transpiler.preset_passmanagers import (
    generate_preset_pass_manager 
)
pm_logs = generate_preset_pass_manager(
    optimization_level=3, 
    basis_gates=['rz','x','sx','cx']
)
\end{lstlisting}
We used the base compiler to lower a MiGS circuit to a LoGS circuit, and optimized the LoGS circuit at the optimization level 3 (the highest).  We did not have to extend or augment it further.

Similarly, we created the base compiler for the MiGS, which includes the CCX and CCRZ gates as the two-qubit-controlled gates, as follows. 
\begin{lstlisting}[language=Python, basicstyle=\ttfamily\small, frame=single]
pm_migs = generate_preset_pass_manager(
    optimization_level=3,     
    basis_gates=['rz','x','h','cx',
                  'crz','ccrz','ccx']
)
\end{lstlisting}
We lowered a HiGS circuit to a MiGS circuit by hand-writing the code to do the {\it vchain} decomposition. In addition, we augmented the base compiler to 
enable the rewriting of a circuit to the equivalence relation in Figure~\ref{fig:ccx2cx} by creating an appropriate instance of the {\tt TemplateOptimiztion}\footnote{in the {\tt qiskit.transpiler.passes module.}} class~\cite{iten2022template}. Note that the MiGS does not include the RCCX gate. It is encapsulated in the implementation. We implemented the replacement described in Section~\ref{sec:ccx2rccx}  by utilizing the {\tt substitute\_node\_with\_dag} method of the {\tt DAGCircuit}\footnote{in the {\tt qiskit.dagcircuit} module.} class.

Likewise, we created the base compiler for the HiGS with a larger set of controlled gates for {\tt basis\_gates}.  Similar to the MiGS, we augmented it to enable the rewriting of a circuit to the equivalence relation in Figure~\ref{fig:cx3cx2}.  

Gate commutation, cancellation, and template matching play crucial roles in simplifying circuits in HiGS and MiGS, or in general in MLCO.  For different gate sets we deploy different sets of target gates for commutation and cancellation and different sets of templates for matching. Among these,  template matching is not light-weighted by nature.  Furthermore, we should run it repeatedly to reach the fixed point, since matching the succeeding template often creates new opportunities for matching the preceding template.

While in this paper we focused on the optimization quality (how fast the resulting circuit runs),  the optimization efficiency (how fast the optimizer runs) is as important.   
Considering we will soon have to optimize circuits with hundreds or even thousands of qubits, the efficiency will be by all means of critical importance.

Obviously, the amount of optimization efforts scales with the circuit size often more than linearly. We thus emphasize that MLCO helps a lot in terms of optimization efficiency as well, since it simplifies the circuit upstream and delivers the simplified downstream.

\section{Results}
\label{sec:results}

\subsection{6-qubit PDE circuit}

\begin{table*}
  \caption{Numbers of entangling gates of PDE circuits}
  \label{tab:PDE}
  \begin{tabular}{l|rrr rr rr r|rr}
    \toprule
    \multirow{2}{*}{~~Circuit (Figure No.)} & \multicolumn{8}{c|}{MLCO} & \multicolumn{2}{c}{DETO} \\
         & C5RZ & C4RZ & C3RZ &CCRZ &CCX & RCCX &CRZ &  CX &  (L0:CX) & CX \\  %
    \midrule
    1-step HiGS source (\ref{fig:higs_stair})  & 1 & 1 & 1 & 1 & 0 & 0 & 1 &  30 & 114 & 102  \\
    1-step HiGS simplified (\ref{fig:hsim_stair})  & 1 & 1 & 1 & 1 & 0 & 0 & 1 &  14 & 98 & 98  \\
    1-step MiGS input (\ref{fig:migs_stair})  & 0 & 0 & 0 & 4 & 12 & 0 & 1 &  14 & 104 & 104  \\
    1-step MiGS simplified (\ref{fig:msim_stair})  & 0 & 0 & 0 & 4 & 6 & 0 & 1 &  16 & 70 & 69  \\    
    \midrule
    2-step MiGS composed (\ref{fig:msim_stair2})  & 0 & 0 & 0 & 8 & 12 & 0 & 2 &  32 & 140 & 138  \\
    2-step MiGS simplified (\ref{fig:msim2_stair2})  & 0 & 0 & 0 & 8 & 6 & 0 & 2 & 24 & 96 & 93  \\ 
    2-step MiGS replaced (\ref{fig:msimRT2_stair2})  & 0 & 0 & 0 & 8 & 0 & 6& 2 & 24 & 78 & 75  \\ 
    2-step LoGS target  & 0 & 0 & 0 & 0 & 0 & 0 & 0 & 75 & 75 & 75 \\ 
    \bottomrule
  \end{tabular}
\end{table*}

As explained in Section~\ref{sec:mlco},
we performed MLCO (Multilevel Circuit Optimization) to the 6-qubit stair-typed PDE circuits most successfully.
We counted the number of entangling gates by type for the progressively lowered circuits from HiGS, MiGS to LoGS. Table~\ref{tab:PDE} shows the 
results. 
As seen in the table,  we ended up with the two-step circuit in LoGS with 75 CX gates, or 37.5 CX gates per step.

We also experimented with the DETO approach to optimize the source circuit, and other circuits as well for curiosity. As mentioned earlier, in DETO, the circuit is first decomposed into single- and two-qubit gates, and the resulting circuit is then optimized at the highest optimization level, with the expectation that massive optimization will happen there.   In our experiment, we obtained the thus-optimized circuits with {\tt pm\_logs} defined in Section~\ref{sec:impl}.  For reference, we also obtained the just-decomposed circuit with a similar instance created with the same gate set but with optimization level 0.  The numbers of CX gates of the just-decomposed and thus-optimized circuits are shown under the DETO columns, (L0:CX) and CX, respectively.

We can make two observations. First, the CX gate number for the circuit is $102$ with DETO, while it is $37.5$ with MLCO.  Thus, our approach achieved 64\% reduction in the number of CX gates.  Second, looking across the circuits, we do not see much difference between the numbers in the just-decomposed circuit and the thus-optimized circuit in DETO, 10.5\% reduction for the source circuit and less than 0.4\% for the other circuits.  Thus, as long as the circuits under our study are concerned, DETO does not live up to the expectation.
Third, we observe a substantial difference between the numbers of CX gates in DETO for 1-step MiGS input and simplified. Simillar, for 2-step MiGS composed and simplified and for 2-step MiGS simplified and replaced. Clearly, these demonstrate the effectiveness of simplification at higher levels.

\subsection{$n$-qubit PDE circuit}

We now consider the $n$-qubit PDE circuit\footnote{In Section~\ref{sec:PDE} $n$ means the number of qubits for discretization, while in this section $n$ means the total number of qubits in the source circuit.} to compare the eventual numbers of CX gates by MLCO and DETO.
While we can hardly derive some formula for them, we can do so for the numbers of the entangling gates by type for the 1-step HiGS source and 2-step MiGS simplified circuits. We can then calculate the numbers of CX gates for the just-decomposed for these circuits, which will serve as the upper bounds of the eventual numbers mentioned above.

The $n$-qubit one-step HiGS source circuit will include one C$k$RZ gate for $1 \le k \le (n-1)$ and $ (2  \sum_{j=1}^{n-1} j)$ CX gates\footnote{There are two wings for each of the C$k$RZ gates, and $k$ CX gates for each wing.}.
We assume the algorithm ~\cite[Theorem 3] {vale2023circuit} which allows C$k$RZ ($k \ge 2) $ to be realized with {\it at most} $(16k-24)$ CX gates because this algorithm is used in Qiskit as the standard decomposition algorithm for multi-controlled rotation gates. We see that Qiskit realizes CRZ, CCRZ, C3RZ, C4RZ, C5RZ, C6RZ, C7RZ and CR$k$Z ($k \ge 8$) with 2, 4, 14, 24, 40, 56, 80, and $(16k-24)$ CX gates, respectively, a simple calculation shows that there are  a total of $(9n^2-33n-36)$ CX gates in the corresponding just-decomposed of the $n$-qubit PDE circuit (for $n \ge 8$).

Similarly, 
the $n$-qubit {\it two-step} MiGS simplified circuit 
contains
$2$ CRZ, $2(n-2)$ CCRZ, $2(n-3)$ RCCX and $2(3n-6)$ CX gates.
Considering that CRZ, CCRZ and RCCX gates are realized with 2, 4, and 3 CX gates, respectively,
a simple calculation shows that there are a total of $2(10n-21)$ CX gates in the corresponding {\it two-step} just-decomposed circuit, 
meaning $(10n-21)$ CX gates {\it per step}.
Compared to DETO, we achieved the {\it quadratic reduction} in the upper bound with MLCO.   

It should be emphasized that each of the C$k$RZ ($3 \le k \le n-1$) in the $n$-qubit source circuit is reduced to just one CCRZ and one CCX in the MiGS simplified circuit. In other words, if you optimize the 100-qubit PDE source circuit by MLCO, the RZ gate controlled by 99 qubits is  reduced to one CCRZ and one CCX!

\begin{figure}[h]
 	\centering
	\includegraphics[width=\linewidth]{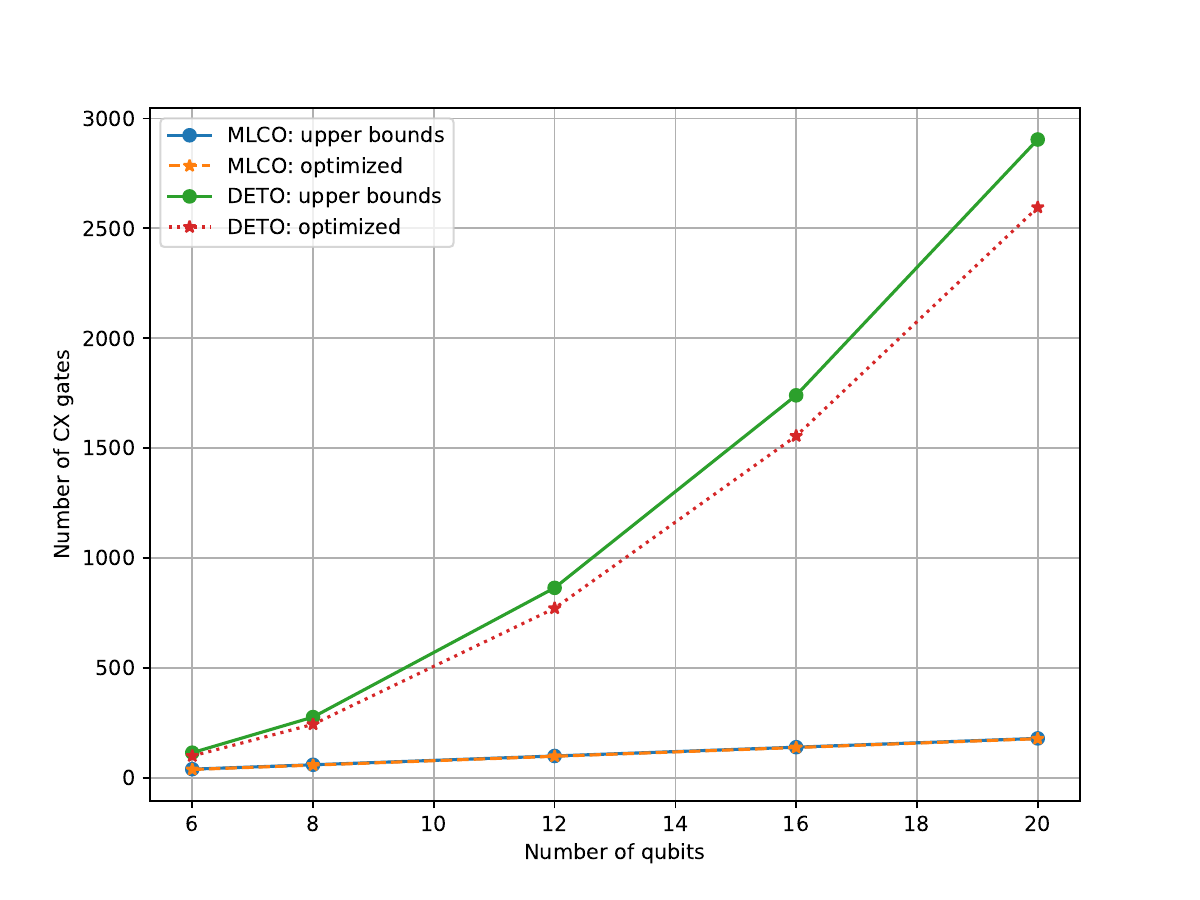}
	\caption{The numbers of CX gates of the optimized circuits by MLCO and DETO of PDE circuits for $n=6,8,12,16$, and $20$, together with the estimated upper bounds (using $114$ for $n=6$).  While the numbers of CX gates increase quadratically with DETO both for the upper bounds and the optimized, they increase linearly with MLCO. Our proposed optimization thus achieved quadratic reduction.}
\label{fig:graph}
\end{figure}

We now empirically optimize different sizes of PDE circuits
by MLCO and DETO,
obtaining the actual (or eventual) numbers of CX gates in the resulting circuits.
Figure~\ref{fig:graph} shows the results together with the upper bounds given by the above formulae. Observing that the actual numbers are very close to the upper bounds, we claim that, with MLCO, we achieved the quadratic reduction in the number of CX gates.

On the other hand, our approach requires $(n-3)$ auxiliary qubits. An $n$-qubit HiGS circuit is lowered into a $(2n-3)$-qubit MiGS circuit. We believe that this is a reasonable tradeoff between the number of qubits and the number of CX gates required, considering the number of qubits and the error rates of CX gates of today's noisy devices.

\section{Related Work}

There exist research efforts with a motivation similar to ours, that is, "do circuit optimization at a high level".
Van Den Berg and Temme~\cite{van2020circuit} introduced a method that partitions Pauli operators into mutually commuting clusters and applies simultaneous diagonalization, effectively reducing both the number of CX gates and the circuit depth in Hamiltonian simulations.
Li et al.~\cite{li2022paulihedral} proposed Paulihedral, 
defining a new IR (Intermediate Representation), called Pauli IR, to represent the quantum simulation kernels at the Pauli string level.  They use the high-level information to capture and create the opportunities of gate cancellation, also exploiting the algorithmic flexibility in synthesis to increase gate cancellation.   
Liu et al.~\cite{liu2024practical} 
applies Paulihedral for $k$-local terms and proposes a template matching algorithm for two-local terms for a practical circuit optimization of Hamiltonian simulation.
Obviously, MLCO pursues circuit optimization at completely different levels from the Pauli strings or operators level. Actually, if we represent the Hamiltonian $h_j$ in the Pauli operators, it ends up with the exponential number of Pauli strings.

There has been a large amount of research on decomposing multi-controlled gates.
For instance, regarding controlled gates, 
some cope with
$U(2)$~\cite{barenco1995elementary,iten2016quantum,rosa2025optimizing,zindorf2024efficient}, 
others deal with  
$SU(2)$~\cite{barenco1995elementary,iten2016quantum,vale2023circuit,rosa2025optimizing,zindorf2024efficient}, 
and still others focus on Pauli gates~\cite{barenco1995elementary,selinger2013quantum,jones2013low,iten2016quantum,maslov2016advantages,he2017decompositions,rosa2025optimizing,zindorf2024efficient}.
Regarding target metrics, 
some attempt to decompose multi-controlled gates with as few CX gates as possible
~\cite{barenco1995elementary,iten2016quantum,vale2023circuit,rosa2025optimizing}, others do so as few T gates ~\cite{selinger2013quantum,jones2013low}  
and still others do so as few of both~\cite{maslov2016advantages,he2017decompositions,zindorf2024efficient}.
In general, more efficient decompositions are made possible with auxiliary qubits ~\cite{barenco1995elementary,he2017decompositions,zindorf2024efficient}.
More recently, qubit-connectivity-aware decompositions are actively pursued, for instance, for LNN (Linear Nearest Neighbor)~\cite{mottonen12006decompositions,chakrabarti2007nearest,miller2011elementary,saeedi2011synthesis,cheng2018mapping,li2023quantum}.
All of these primarily focus on individual multi-controlled gates.
In this paper, we do not propose any new decomposition algorithm. Rather, we do propose to choose a good decomposition algorithm based on a high-level circuit structure made visible by MLCO. Thus, all the above algorithms are in our decomposition arsenal.

\section{Conclusion}

We explored MLCO (Multilevel Circuit Optimization), 
using the PDE (Partial Differential Equations) circuit dense with multi-controlled gates as a case study.
We demonstrated its  effectiveness, 
having successfully achieved quadratic reduction
in the number of CX gates. The key is to deploy multiple gate sets and progressively lower a higher-level circuit to a lower-level circuit.   
Having a clear picture of a circuit at a high level provided us insights to how to simplify and optimize the circuit.
In particular, we learned that we could create a circuit structure
to cause massive cancellation of entangling gates
by selecting the right algorithm to decompose multi-controlled gates
and by composing equivalent but different circuits.

There are many interesting future research directions. An immediate one is to apply MLCO to more circuits
derived from real problems with business and science values.
The more experience, the better understanding we have of what intermediate gate sets to deploy and what optimizations to perform at each gate set. 
Another direction would be to explore connection between circuit optimization and physics.
Recall that we started with physics (Hamiltonian) to build the source circuit, and that we eventually transformed it to the target circuit, performing lots of optimizations.  An interesting question would then be what interpretations are possible in the physics world for what  happened in the circuit world.

\begin{acks}
This work is supported by MEXT Quantum Leap Flagship Program Grant Number JPMXS0118067285 and JP- MXS0120319794, and JSPS KAKENHI Grant Number 20H05966. We acknowledge the use of IBM Quantum services for this work. The views expressed are those of the authors, and do not reflect the official policy or position of IBM or the IBM Quantum team. We would like to thank Kenji Sugisaki and Jumpei Kato of Keio University and Matthew Treinish of IBM Quantum for the fruitful discussions on circuit optimization.

\end{acks}

\bibliographystyle{ACM-Reference-Format}
\bibliography{reference}

\end{document}